\documentstyle[prb,aps]{revtex}
\begin{document}
\draft

\title{The Role of Nonequilibrium Dynamical Screening \\
  in Carrier Thermalization}

\author{Girish S. Setlur and  Yia-Chung Chang}
\address{Department of Physics and Materials Research Laboratory 
\\ University of Illinois at Urbana-Champaign, Urbana, Illinois 61801}

\date{\today}
\maketitle


\begin{abstract}

 We investigate the role played by nonequilibrium dynamical screening
 in the thermalization of carriers in a simplified two-component
 two-band model of a semiconductor. The main feature of our
 approach is the theoretically sound treatment of collisions.
 We abandon Fermi's Golden rule in favor of the 
 Schwinger-Bakshi-Mahantappa-Keldysh (nonequilibrium field theoretic)
 formalism as the former is applicable
 only in the long-time regime. We also introduce the
 concept of nonequilibrium dynamical screening.
 The dephasing of excitonic quantum beats as a result of carrier-carrier
 scattering is brought out. At low densities it is found that the dephasing
 times due to carrier-carrier scattering is in picoseconds and not
 femtoseconds, in agreement with experiments.
 The polarization
 dephasing rates are computed as a function of the excited carrier density
 and it is found that the dephasing rate for carrier-carrier scattering
 is proportional to the carrier density at ultralow densities. The scaling
 relation is sublinear at higher densities, which enables a comparison
 with experiment. 

\end{abstract}

\section{Introduction}

 The relaxation of hot electron distributions via emission of LO optical
 phonons has been studied by a number of authors. However, the
 role of carrier-carrier scattering has received less attention for two main
 reasons. The theory of screening involved in the proper description of
 carrier-carrier scattering is poorly understood. Any reasonable attempt
 to include carrier-carrier scattering is computationally intensive.
 Secondly, the perception that carrier-carrier scattering is
 significant only for large carrier densities has resulted in few attempts at
 accurately modeling the phenomenon. In particular, authors in 
 Refs.~\onlinecite{1,2,3,4}
 have conducted theoretical studies of the relaxation of hot electrons
 when the emission of LO optical phonons is the dominant mechanism.
 The authors in Ref.~\onlinecite{5,6,7} have attempted to investigate the
 role of carrier-carrier
 scattering in momentum and energy relaxation. On the experimental side,
 authors in Refs.~\onlinecite{8,9,10,11,12,13,14,15} have investigated the 
relaxation of hot  carriers in
 bulk GaAs as well as quantum well structures such as GaAs/AlGaAs.
 A number of experiments have been performed in the recent years
 on the density dependence of the scattering rate of free carriers
 which provide information about the role of carrier-carrier 
 scattering. Prominent among them are the experiments by authors
 in Refs. \onlinecite{5} and \onlinecite{16,17,18,19,20,21,22}. 
The use of nonequlibrium Green function techniques
 in the study of relaxation phenomena is not widespread. However, attempts
 have been made in the recent past. Notable among those are the study of
 memory effects due to relaxation by phonons by authors in 
Ref.~\onlinecite{34}
 and an attempt by authors in Ref.~\onlinecite{35} 
 to model the nonequlibrium aspects of screening of
 the Coulomb interaction, similar to the approach outlined here. 

        It has been pointed out in Ref.~\onlinecite{7} that changes in the scattering rates
 with carrier density are solely due to coulomb scattering of carriers.
 Therefore an investigation of the density dependence of
 carrier-carrier scattering should provide a good testing ground for
 theories of coulomb scattering. The authors in Ref. \onlinecite{5,6,7} have
 developed a theory of carrier-carrier scattering using the quasi-classical
 Boltzmann equation coupled with Fermi's Golden rule for evaluating the
 the collision terms. Objections to the use of Fermi's Golden rule
 may be raised, however. Firstly, Fermi's Golden rule is derived from
 time-dependent perturbation theory as an asymptotic time approximation
 which translates into a strict energy conservation rule.
 Quantitatively, the time resolution
 of a typical photon echo measurement\cite{16} is of the order of 
 $\Delta t \sim 10 fs. $. This translates to an energy uncertainty of an 
order of $ \Delta E \sim 80 meV $. This is substantial when compared
 to the plasmon energy at densities  of the order of 
$ \rho_0 \sim 2 \times 10^{16}/cm^{3} $
or the LO phonon energy which is $ \omega_{LO} = 36 meV $.
 At these densities the
 plasmon energy is $ \omega_{p} \sim 6.2 meV $. 
 We therefore abandon the familiar
 rule in favor of a more elaborate explicitly nonequilibrium field-theoretic
 formulation which we shall discuss subsequently. There is at least one
 other significant conceptual drawback in the present theories of Coulomb
 scattering in carrier thermalization. We are primarily interested in the
 dynamics that occur within 10 ps. of the switching on of the external
 field. Within this time scale, the system is in a highly
 nonequilibrium state. Therefore it does not, in particular, make sense 
 to use the naive generalisation of the equilibrium dynamically
 screened coulomb interaction which essentially involves replacing
 in the formula for the RPA dielectric function, the equilibrium
 carrier densities by nonequilibrium ones. This is because when
 time translation invariance is broken as is the case here, the
 RPA (random phase phase approximation)
 as it is commonly understood, is no longer valid as it is not possible to
 transform to the frequency domain. 
 Our field-theoretic formulation involves the use of
 Schwinger-Bakshi-Mahantappa-Keldysh-Kadanoff-Baym\cite{23,24,25,26}
   approach in deriving the collision terms (henceforth the phrase
 'collision terms' will refer to the right side of the 
 Boltzmann equation. A good introduction to
 this approach may be found in Ref.~\onlinecite{26}.) We assume that the
 collision terms in the Boltzmann equation can be
 expanded in a power series of the external (time-dependent)
 vector potential. This approach is simliar to linear
 (or quadratic) response theory.
 The main difference being that we solve the full Boltzmann equation
 numerically after the collision terms have been obtained so that the
 densities and currents no longer depend only on finite powers of the 
 external fields. We also obtain explicit formulas for the screened coulomb
 interaction in terms of the external fields. These formulas
 clearly indicate the nonequilibrium nature of the dynamical screening. 
 This screened potential also produces a finite collision cross-section
 in both two and three space dimensions.
 
\section{Semiconductor Bloch Equations} 

 We consider a three dimensional, two component, undoped, electron-hole system 
 interacting with a transverse classical electromagnetic field along with
 a source for the electromagnetic field in addition to the usual
 coulombic carrier-carrier interaction. 
 It is important to include
 other terms such as electron-phonon interaction as this is the dominant
 mechanism of energy relaxation. Impurity scattering is relatively less
 important for the ultra-pure sample that we consider. 
 Further, it
 has been argued in the literature that only carrier-carrier scattering
 contributes significantly to the density dependence of the scattering rates.
 Intuitively, this is understandable as the electrons can independently
 emit phonons at a fixed rate irrespective of how many electrons are
 participating in the process. By contrast, the carrier-carrier scattering rate
 depends on how many carriers there are to scatter off from. We shall
 verify these expectations rigorously later on.
	 
 We assume that the system is in thermodynamic
 equilibrium at temperature $ T = 0 $ before time $ t = 0 $
 (we use natural units throughout, $ c = \hbar = 1 $).
 The fields are assumed to have been turned on at $ t = 0 $ and
 last for a duration of $ \tau_{X} = 250fs. $.  The excitation energy
 coincides with the band gap $ \omega_{X} = E_{g}-E_b $, where $E_g$ and
$E_b$ are the band gap and exciton binding energy, respectively.
 Fig.1 is a schematic illustration of the conditions under which the
 simulations are performed.
 The semiconductor Bloch equations (SBE for short) (see for example 
Ref.~\onlinecite{32})
 are nothing but the equations of motion for the equal time
 component of the Green functions of the system.
 The equations in the form given below were first derived in 
Refs.~\onlinecite{27,28,29,30} (Our notation is slightly different though).
 The Green functions in the equation below are proportional to the
 inter-band polarization and the electron and hole density distributions. 

\[
g_{hh}({\bf{k}}t)=i\langle \psi_{h}^{\dagger}({\bf{k}}t)
\psi_{h}({\bf{k}}t) \rangle = i \mbox{ } f({\bf{k}},t)
\]

\[
g_{he}({\bf{k}}t)=i\langle \psi_{h}({\bf{k}}t)\psi_{e}({\bf{k}}t) \rangle 
\]

\[
i\frac{\partial}{\partial t}g_{hh}({\bf{k}}t)
 = 2Re(\Omega({\bf{k}}t)g^{*}_{he}({\bf{k}}t)) + R_{hh}({\bf{k}}t)
\]

\[
i\frac{\partial}{\partial t}g_{he}({\bf{k}}t)
= -\Omega({\bf{k}}t)(i-2g_{hh}({\bf{k}}t))
 + (\epsilon_{h}({\bf{k}})+\epsilon_{c}({\bf{k}})-2\Sigma({\bf{k}}t))
g_{he}({\bf{k}}t) + R_{he}({\bf{k}}t)
\]

\[
\Omega({\bf{k}}t) = {\frac{e}{mc}}{\bf{A}}_{ext}(t).{\bf{p_{vc}}} 
- i\sum_{{\bf{k^{'}}}}v_{{\bf{k-k^{'}}}}g_{he}({\bf{k^{'}}}t)
\]

\begin{equation}
\Sigma({\bf{k}}t) = -i\sum_{{\bf{k^{'}}}}v_{{\bf{k-k^{'}}}}
g_{hh}({\bf{k^{'}}}t)
\end{equation}
 Here we have tacitly assumed that $ g_{hh}({\bf{k}}t) $ is equal to
  $ g_{ee}({\bf{k}}t) $. This is an approximation that we justify
 rigorously later.
 The collision terms $ R_{hh}({\bf{k}}t) $ and $ R_{he}({\bf{k}}t) $
 are in general, extremely complicated functions of the full 
 Green-functions of the system. (Collisionless or Hartree-Fock(HF)
 approximation
 therefore means setting $  R_{hh}({\bf{k}}t) $ and $ R_{he}({\bf{k}}t) $
 to zero.). Perhaps the simplest approximation for these terms would be to use
 constant relaxation rates:
 $ R_{hh}({\bf{k}}t) = -i \mbox{ } g_{hh}({\bf{k}}t)/T_{1} $,
 $ R_{he}({\bf{k}}t) = -i \mbox{ } g_{he}({\bf{k}}t)/T_{2} $.
 The polarization dephasing rate is defined as $ \gamma = 1/T_{2} $.
 and $ 1/T_{1} $ is the population relaxation rate.
 These aprroximations are too crude for most purposes unless one is
 interested exclusively in studying coherent phenomena. 
The approximation
 for $ R_{hh}({\bf{k}}t) $ in terms of a constant population relaxation
 rate suggests that the total kinetic energy of the holes decays with
 the same rate as the
 total number of holes. We know that this is not true, because when phonons
 are present the kinetic energy relaxation takes place on a time scale
 of picoseconds whereas population relaxation through recombination
 occurs on a time scale of nanoseconds. Therefore the approximation
 for $ R_{hh}({\bf{k}}t) $ in terms of a constant relaxation rate is
 particularly bad.
 A more sophisticated approach would involve the use of 
 Fermi's Golden rule to evaluate the collision terms. Here, the relaxation
 terms have a complicated dependence on the polarization and densities.
 However, the rates $ R_{hh}({\bf{k}}t) $ and $ R_{he}({\bf{k}}t) $
 still depend on the polarization and densities at time $ t $ and not on
 all previous times. Higher in the hierarchy of theoretical sophistication
 is the nonequilibrium Green function approach.
 Here, the collision terms depend on the full Green function of the system and 
 therefore effectively, on the polarization and densities at all times previous 
 to time $ t $ at which the collision rates are evaluated. 
 Such a depndence of the collision terms on the history of the system is
 known as the memory effect. Each approach has its pros and cons. 
 While the method of using constant relaxation rates has in its favor,
 simplicity, it contains phenomenological parameters that have to be fitted
 from experiment. In fact the phenomenon of the photon echo\cite{16} has been
 succesfully used to measure the polarization dephasing rate
 ($ \frac{1}{T_{2}} $). The use of
 Fermi's Golden rule works well for studying relaxation via phonons but is 
 ambiguous when applied to carrier-carrier scattering (for reasons mentioned
 in the introduction and for reasons that will become clearer later.). 
 In light of these arguments we take an
 alternative route. The collision terms in the Boltzmann
 equation occur as the product of the collision self-energy and the 
 full Green function. The collision self-energy in turn
 depends on the full Green function.
 When the external fields are absent the collision terms are identically
 zero. Therefore, it is sensible to expect a power series expansion
 for the collision terms in powers of the external fields.
 In order to simplify matters further, we assume that for the purpose
 of evaluating these terms, it is sufficient to retain only the coupling
 of the carriers to the electromagnetic field, as the latter is responsible
 for generating all the dynamics. 
 Therefore, the zeroth term in this expansion is just the
 collision self-energy evaluated using the free Green function
 multiplied by the free Green function which turns out to be identically zero.
 The next term would involve the first power of the total
 vector potential multiplied by a linear response term which we
 evaluate. A more precise quantitative meaning of this approximation
 will be given in the next section. The screened fock self-energy 
 ( the GW approximation as it is sometimes called )\cite{26} involves the
 evaluation of the screened coulomb interaction. To this end, we
 introduce the concept of nonequilibrium dynamical screening.

\subsection{Nonequilibrium Dynamical Screening}

 The RPA is obtained by linearizing the Hartree
 equations of motion for the Green function in the external field.
 Alternatively, it may be
 obtained by replacing the full Green function by the Hartree Green function
 in the shielded potential approximation as discussed in Ref.~\onlinecite{26}.
 This is exactly what we do, except that in our case the Hartree Green function
 lacks time translation invariance and therefore it is not possible to
 solve the resulting integral equation for the sheilded potential
 by transforming to the frequency domain. This will force us to seek an
 expansion in terms of a small parameter. 

 In order to derive the relevant formulas for the screened coulomb
 interaction, we proceed as follows. The equation of motion for the 
 mean-field (Hartree) Green function of the system, in the presence
 of an external classical electromagnetic field may be written as:
\begin{equation}
(i\frac{\partial}{\partial t_{1}} - H_{0}(1))G(1,1^{'}) = 
\delta(1-1^{'}) + \frac{e}{mc}{\bf{A}}(1).{\bf{p^{T}_{1}}}
\mbox{ }M\mbox{ }G(1,1^{'}) 
 + \frac{e^{2}}{2mc^{2}}|{\bf{A}}(1)|^{2}G(1,1^{'}).
\end{equation}
Here and henceforth, we use the abbreviation $ 1 = ({\bf{x_{1}}}t_{1})$.
This Green function will be used as the starting point for derivation of the
formulas for the screened coulomb interaction. Note here that the
 Hartree-Coulomb terms are absent because of charge neutrality.
 The Green function in the above equation  is defined as
\begin{equation}
G(1,1^{'}) = \left( \begin{array}{clcr}
			G_{ee}(1,1^{'}) & G^{*}_{eh}(1,1^{'}) \\
			G_{he}(1,1^{'}) & G_{hh}(1,1^{'}) 
		    \end{array} \right), 
\end{equation}
and
\begin{equation}
M = \left( \begin{array}{clcr}
                        0 & 1 \\
                        1 & 0
                    \end{array} \right).
\end{equation}
Here the contour-ordered Green functions are defined as,
\[
G_{hh}(1,1^{'}) =
\frac{-i\langle T_{c} S\mbox{ } \psi_{h}(1)
\psi^{\dagger}_{h}(1^{'}) \rangle}{T_{c} S},
\]
\[
G_{ee}(1,1^{'}) = 
\frac{-i\langle T_{c} S\mbox{ } \psi_{e}^{\dagger}(1)
\psi_{e}(1^{'}) \rangle}{T_{c} S},
\]
\[
G^{*}_{eh}(1,1^{'}) =
\frac{-i\langle T_{c} S\mbox{ } \psi_{e}^{\dagger}(1)
\psi_{h}^{\dagger}(1^{'}) \rangle}{T_{c} S},
\]
\begin{equation}
G_{he}(1,1^{'}) =
\frac{-i\langle T_{c} S\mbox{ } \psi_{h}(1)
\psi_{e}(1^{'}) \rangle}{T_{c} S}.
\end{equation}
Here $ S $ is the S-matrix that involves the part of the hamiltonian that 
generates the external field. Using the same methods outlined in
Ref.~\onlinecite{26} the screened coulomb interaction $ v_{S}(1,3) $ satisfies:
\[
v_{S}(1,3) = v(1-3) - i(2S+1)\int\mbox{ }d2 \int\mbox{ }d4\mbox{ }
v_{S}(1,2)F(2,4)v(4-3),
\]
\begin{equation}
F(2,4) = G_{he}(4,2)G^{*}_{eh}(2,4^{+}) + G_{hh}(4,2)G_{hh}(2,4^{+})
+ G_{ee}(4^{+},2)G_{ee}(2,4) + G^{*}_{eh}(4^{+},2)G_{he}(2,4).
\end{equation}
We can now expand $ v_{S}(1,3) $ in powers of the vector potential
 by expanding the Green-function in powers of the
 same using the equation of motion for the mean-field Green function.
 The dimensionless quantity that appears in this expansion is denoted by
 $ \Lambda({\bf{k}}t) $. The expansion in powers of the vector
 potential is an expansion in powers of the dimensionless quantity defined
 below.

\[
\Lambda({\bf{k}},t) =
\int_{0}^{t} dt^{'} \mbox{ } \frac{e}{mc}{\bf{A}}_{ext}(t^{'})
\cdot {\bf{p_{cv}}}exp(-i\epsilon({\bf{k}})t^{'}),
\]
\begin{equation}
\Lambda_{0} = \frac{e}{mc}{\bf{ A_{ext}.p_{cv} } }\tau_{X}.
\end{equation}
 Here $ t $ is some arbitrary time after $ t = 0 $ and $ A_{ext} $ is the
 amplitude of $ {\bf{A}}_{ext}(t) $ and $ \tau_{X} = 250fs. $ is the
 pump duration. 
 We require the magnitude of $ \Lambda({\bf{k}},t) $ quantity to be small
 compared with unity for all values of
 $ \epsilon({\bf{k}}) = \frac{k^{2}}{2\mu} + E_{g} $ which is the
 interband transition energy at wavevector $ {\bf{k}} $. The expansion in
 powers of $ \Lambda $ is synonymous with working at low-densities.
 The leading terms in the expansion for the Green functions can be derived by 
 expanding the Green function in powers of the vector potential. 
\begin{equation}
G(1,1^{'}) = G_{0}(1,1^{'}) +
\int d3 \mbox{ } (\frac{\delta G(1,1^{'})}{\delta A(3)})_{0}A(3)
+ \frac{1}{2} \int d3 \mbox{ }\int d4 \mbox{ }
(\frac{\delta^{2} G(1,1^{'})}{\delta A(3) \delta A(3)})_{0}A(3)A(4).
\end{equation}
The derivatives are computed in the usual manner by first inverting 
Eq. (2) to obtain an equation for the inverse $ G^{-1}(1,1^{'}) $.
\begin{equation}
G^{-1}(1,1^{'}) =
 (i\frac{\partial}{\partial t_{1}} - H_{0}(1))\delta(1-1^{'})
- \frac{e}{mc}{\bf{A}}(1).{\bf{p^{T}_{1}}}\mbox{ }M\mbox{ }\delta(1-1^{'})
- \frac{e^{2}}{2mc^{2}}|{\bf{A}}(1)|^{2}\delta(1-1^{'})
\end{equation}
and then by using relations such as,
\begin{equation}
\delta G(1,1^{'}) = -\int d3 \mbox{ }
 \int d4 \mbox{ }G(1,3) (\delta G^{-1}(3,4)) G(4,1^{'}).
\end{equation}
Such an approach leads us to expressions for the leading terms
 in the expansion for the Green functions of the system.
\[
G^{<}_{ee}(1,1^{'}) = i\sum_{{\bf{k}}}
\phi_{{c}{\bf{k}}}({\bf{x_{1}}})\phi^{*}_{{c}{\bf{k}}}({\bf{x_{1^{'}}}})
exp(i \epsilon_{c}({\bf{k}})(t_{1}-t_{1^{'}})),
\]
\[
G^{>}_{ee}(1,1^{'}) = -i\sum_{{\bf{k}}}
\phi_{{c}{\bf{k}}}({\bf{x_{1}}})\phi^{*}_{{c}{\bf{k}}}({\bf{x_{1^{'}}}})
exp(i \epsilon_{c}({\bf{k}})(t_{1}-t_{1^{'}})  )
\Lambda({\bf{k}}t_{1})\Lambda^{*}({\bf{k}}t_{1^{'}}),
\]
\[
G_{eh}^{*>}(1,1^{'}) = -\sum_{{\bf{k}}}\phi_{{c}{\bf{k}}}({\bf{x_{1}}})
\phi^{*}_{{v}{\bf{k}}}({\bf{x_{1^{'}}}})
exp(i\epsilon_{c}({\bf{k}})t_{1})exp(i\epsilon_{h}({\bf{k}})t_{1^{'}})
\Lambda({\bf{k}}t_{1}),
\]
\[
G_{eh}^{*<}(1,1^{'}) = -\sum_{{\bf{k}}}\phi_{{c}{\bf{k}}}({\bf{x_{1}}})
\phi^{*}_{{v}{\bf{k}}}({\bf{x_{1^{'}}}})
exp(i\epsilon_{c}({\bf{k}})t_{1})exp(i\epsilon_{h}({\bf{k}})t_{1^{'}})
\Lambda({\bf{k}}t_{1^{'}}),
\]
\[
G_{he}^{>}(1,1^{'}) = \sum_{{\bf{k}}}\phi_{{v}{\bf{k}}}({\bf{x_{1}}})
\phi^{*}_{{c}{\bf{k}}}({\bf{x_{1^{'}}}})
exp(-i\epsilon_{h}({\bf{k}})t_{1})exp(-i\epsilon_{c}({\bf{k}})t_{1^{'}})
\Lambda^{*}({\bf{k}}t_{1^{'}}),
\]
\[
G_{he}^{<}(1,1^{'}) = \sum_{{\bf{k}}}\phi_{{v}{\bf{k}}}({\bf{x_{1}}})
\phi^{*}_{{c}{\bf{k}}}({\bf{x_{1^{'}}}})
exp(-i\epsilon_{h}({\bf{k}})t_{1})exp(-i\epsilon_{c}({\bf{k}})t_{1^{'}})
\Lambda^{*}({\bf{k}}t_{1}),
\]
\[
G^{<}_{hh}(1,1^{'}) = i\sum_{{\bf{k}}}
\phi_{{v}{\bf{k}}}({\bf{x_{1}}})\phi^{*}_{{v}{\bf{k}}}({\bf{x_{1^{'}}}})
exp(-i \epsilon_{h}({\bf{k}})(t_{1}-t_{1^{'}})  )
\Lambda({\bf{k}}t_{1^{'}})\Lambda^{*}({\bf{k}}t_{1}),
\]
\begin{equation}
G^{>}_{hh}(1,1^{'}) = -i\sum_{{\bf{k}}}
\phi_{{v}{\bf{k}}}({\bf{x_{1}}})\phi^{*}_{{v}{\bf{k}}}({\bf{x_{1^{'}}}})
exp(-i \epsilon_{h}({\bf{k}})(t_{1}-t_{1^{'}})).
\end{equation}
 From the above formulas it is obvious that the leading contribution to the
 collision part of the screened coulomb interaction depends
 quadratically on the vector potential.
\begin{equation}
v_{S}(1,2) 
=\int \mbox{ }d3 \mbox{ }d4 \frac{1}{2}
(\frac{\delta^{2} \mbox{ }v_{S}(1,2)}{\delta A(3) \delta A(4)})_{0}
A(3)A(4)
\end{equation}
 Assuming space translational invariance (in the one conduction band and
 one valence band model), the (fourier transform of the collision part of)
 the screened coulomb interaction is 

\[
v^{>}_{{\bf{q}}}(t_{1}, t_{2})
 = -i\frac{1}{\Omega}(2S+1)v_{{\bf{q}}}^{2}
\sum_{{\bf{k}}}
\{-F_{1}({\bf{k,q}};t_{1},t_{2})\Lambda({\bf{k+q}};t_{1})
\Lambda^{*}({\bf{k}};t_{2})
\]
\[
- F_{1}^{*}({\bf{k,q}};t_{2},t_{1})\Lambda({\bf{k}};t_{1})
\Lambda^{*}({\bf{k+q}};t_{2})
\]
\[
+ F_{h}
({\bf{k,q}}; t_{2}-t_{1})\Lambda({\bf{k}};t_{1})\Lambda^{*}({\bf{k}};t_{2})
\]
\begin{equation}
+ F_{e}({\bf{k,q}}; t_{2}-t_{1})\Lambda({\bf{k}};t_{1})
\Lambda^{*}({\bf{k}};t_{2}),
\}
\end{equation}
\begin{equation}
v^{<}_{{\bf{q}}}(t_{1}, t_{2})
 = -(v^{>}_{{\bf{-q}}}(t_{1}, t_{2}))^{*},
\end{equation}
\[
F_{1}({\bf{k,q}};t_{1},t_{2}) = 
exp(i(\epsilon_{h}({\bf{k+q}})-\epsilon_{h}({\bf{k}}))t_{2})
exp(i(\epsilon_{c}({\bf{k+q}})-\epsilon_{c}({\bf{k}}))t_{1}),
\]
\[
F_{h}({\bf{k,q}}; t_{2}-t_{1}) = 
exp(i(\epsilon_{h}({\bf{k+q}})-\epsilon_{h}({\bf{k}}))(t_{2}-t_{1})),
\]
\[
F_{e}({\bf{k,q}}; t_{2}-t_{1}) =
exp(i(\epsilon_{c}({\bf{k-q}})-\epsilon_{c}({\bf{k}}))(t_{2}-t_{1})),
\]
 $ v_{\bf{q}} = \frac{2 \pi e^{2}}{\epsilon_{0} q} $ in two dimensions,
 and $ v_{\bf{q}} = \frac{4 \pi e^{2}}{\epsilon_{0} q^{2}} $ in three
 dimensions. The spin $ S = 1/2 $ for fermions.
 $ \epsilon_{c}({\bf{k}}) = \frac{k^{2}}{2m^{*}_{e}} + E_{g} $ and
  $ \epsilon_{h}({\bf{k}}) = \frac{k^{2}}{2m^{*}_{h}} $ and the effective
 masses are $ m^{*}_{e} = 0.067\mbox{ }m_{e} $ and
 $ m^{*}_{h} = 0.5\mbox{ }m_{e} $ and $ \epsilon_{0} = 12.4 $.

 From the above formulas for the screened coulomb interaction it is
 clear that for small $ q $ the screened coulomb interaction 
 $ v^{>}_{{\bf{q}}}(t_{1},t_{2}) $ is proportional to $ \frac{1}{q^{2}} $
 in three space dimensions and independent of $ q $ in two space
 dimensions. The sum over all  $ q $'s has a $ q^{2} $ in the measure in
 3D which cancels the divergence coming from $ \frac{1}{q^{2}} $.
 In 2D the measure has a $ q $ which mutiplies a constant and 
 vanishes for small $ q $.
 Therefore, in both these cases we obtain a finite
 collision cross-section: an important and essential feature of
 any theory of screening. The above formulas differ from the naive
 generalisation of the RPA in some respects.
 Firstly, the traditional RPA is invoked in a slightly different context.
 Here, the electrons are treated as quantum objects immersed in a
 uniform positive background which does not participate in the
 dynamics. The uniform positive background ensures charge neutrality.
 The RPA then is just the linear response of such a system
 to weak external longitudinal (scalar) potentials. 
 In our case, however, the external fields are weak transverse electromagnetic
 fields, but more importantly, the carriers responsible for the screening
 are generated out of the vacuum by these weak external fields themselves.
 So it is important to use a nonequlilibrium formalism from the start.
 The connection between the formulas derived
  above and RPA in the steady state limit becomes clear when one realises that
 in the steady state limit the Green functions possess time translation
 invariance and therefore Eq. (6) can be fourier transformed and inverted
 algebraically to yield an expression for the screened coulomb interaction.
 This screened interaction is expressed in terms of the full (steady state)
 Green function of the system. (Physically, the steady state limit can be
 realised by applying an infinitely weak field for an infinitely long time.)
  When the full Green function is replaced
 with the the mean-field Hartree Green function, we recover the familiar RPA.
 If we assume further that the densities involved are small, it is easy
 to see that the collision part of the screened coulomb interaction is
 quadratic in the unscreened
 potential. This is equivalent to iterating the integral equation (Eq. (6))
 once. The steady state limits of Eqs. (13) and (14) are
 therefore identical to the
 low density limit of RPA. However, the RPA (i.e., the replacement of the
 full Green function by the Hartree Green function in Eq. (6)) is known to be
 exact only in the ultra high density limit and its low density limit 
 is not likely to capture all the physics. In light of the complexity of the
 problem we are dealing with here, and all the other
 approximations that we have made, it is reasonable to proceed further,
 after pointing out these pitfalls to the reader.

 Quasistatic screening is the popular choice in the literature\cite{7}. 
But it is well known that this type of screening grossly overestimates
 the importance of screening yielding an exponentially decaying Yukawa type
 of potential in real space.
 The naive generalisation of the RPA
 involves replacing in the formula for the equilibrium RPA, equilibrium
 densities(and polarizations) by nonequilibrium ones. When this is done,
 it not not clear
 what the meaning of the frequency that appears in the dielectric function
 is. Nor is it clear how that screened coulomb interaction should be
 coupled with Fermi's Golden Rule.
 Our approach is more systematic and perhaps
 more appealing. The approach by authors in Ref.~\onlinecite{35} 
comes closest to the one 
 just discussed above. The authors however do not solve the SBE including
 the detailed nonequlibrium screening that they introduce.

\subsection{Collision Terms due to Screening of the Coulomb Interaction}
 The collision terms are evaluated using the same scheme that was used in 
 the evaluation of the screened coulomb interaction.
 $ R_{he}({\bf{k}}t) $ is expanded in powers of $ \Lambda({\bf{k}}t) $.
 The collision self-energy is 
 $ \Sigma_{c}(1,1^{'}) = iv_{S}(1,1^{'})G(1,1^{'}) $
 Defining
\[
 R(1,1^{'}) = \left( \begin{array}{clcr}
                        R_{ee}(1,1^{'}) & R^{*}_{eh}(1,1^{'}) \\
                        R_{he}(1,1^{'}) & R_{hh}(1,1^{'})
                    \end{array} \right) 
 = \int d2 \mbox{ }\Sigma_{c}(1,2)G(2,1^{'}) 
 - \int d2 \mbox{ }G(1,2)\Sigma_{c}(2,1^{'}),
\]
the relevant component of $ R(1,1^{'}) $ namely, 
\[
 R_{he}({\bf{x_{1}}}t_{1},{\bf{x_{1^{'}}}}t_{1^{+}}) =
 \sum_{{\bf{k}}}\phi_{{v}{\bf{k}}}({\bf{x_{1}}})
\phi^{*}_{{c}{\bf{k}}}({\bf{x_{1^{'}}}}) R_{he}({\bf{k}}t_{1}) 
\]
yields the necessary collision term. An expansion of $ R(1,1^{'}) $
in powers of $ \Lambda $  implies that the collision term 
$ R_{he}({\bf{k}}t) $ is proportional to $ \Lambda^{3} $.
\[
R_{he}({\bf{k}}t) = 
{\frac{1}{V}}\sum_{{\bf{q}}}\int_{0}^{t}dt_{2}\mbox{ }
[U_{1}({\bf{k,q}};t,t_{2})\Lambda^{*}({\bf{k-q}},t_{2})
- U_{2}({\bf{k,q}};t,t_{2})\Lambda^{*}({\bf{k}},t_{2})],
\]
\[
U_{1}({\bf{k,q}};t,t_{2}) = v^{>}_{{\bf{q}}}(t,t_{2})
exp(-i(\epsilon_{h}({\bf{k-q}})+\epsilon_{c}({\bf{k}}))t)
exp(-i(\epsilon_{c}({\bf{k-q}})-\epsilon_{c}({\bf{k}}))t_{2})
\]
\[
+ v^{<}_{{\bf{q}}}(t_{2},t)
exp(-i(\epsilon_{h}({\bf{k-q}})-\epsilon_{h}({\bf{k}}))t_{2})
exp(-i(\epsilon_{c}({\bf{k-q}})+\epsilon_{h}({\bf{k}}))t),
\]
\[
U_{2}({\bf{k,q}};t,t_{2}) = v^{>}_{{\bf{q}}}(t,t_{2})
exp(i(\epsilon_{h}({\bf{k-q}})-\epsilon_{h}({\bf{k}}))t_{2})
exp(-i(\epsilon_{h}({\bf{k-q}})+\epsilon_{c}({\bf{k}}))t)
\]
\begin{equation}
+ v^{<}_{{\bf{q}}}(t_{2},t)
exp(-i(\epsilon_{c}({\bf{k}})-\epsilon_{c}({\bf{k-q}}))t_{2})
exp(-i(\epsilon_{h}({\bf{k}})+\epsilon_{c}({\bf{k-q}}))t),
\end{equation}
 and $ R_{hh}({\bf{k}},t) $ and $ R_{ee}({\bf{k}},t) $
 are of the order of $ \Lambda^{4} $ and are therefore neglected as far as 
 coulomb scattering is concerend.

\subsection{Collision Terms due to Phonons}

For treating the electron-phonon collisions we must expand both 
$ R_{hh}({\bf{k}}t) $, $ R_{ee}({\bf{k}}t) $ and $ R_{he}({\bf{k}}t) $
 in powers of $ \Lambda({\bf{k}}t) $. The expansion yields the
 following from for the collision terms.
\[
R_{he}({\bf{k}}t) = -\int_{0}^{t} dt_{2}\mbox{ } 
\frac{2}{V}\sum_{{\bf{q}}}|M_{{\bf{q}}}|^{2}
 \sin(\omega_{LO}(t_{2}-t))
[exp(i ( \epsilon_{c}({\bf{k-q}})-\epsilon_{c}({\bf{k}}) ) t_{2})
\]
\[
exp(-i ( \epsilon_{c}({\bf{k-q}})+\epsilon_{h}({\bf{k}}) ) t )
\Lambda^{*}({{\bf{k}}}t_{2})
\]
\[
 - exp(i ( \epsilon_{h}({\bf{k}})-\epsilon_{h}({\bf{k-q}}) ) t_{2} )
\]
\[
exp(-i ( \epsilon_{c}({\bf{k-q}})+\epsilon_{h}({\bf{k}}) ) t )
\Lambda^{*}({{\bf{k-q}}}t_{2})]
\]
\[
-\int_{0}^{t} dt_{2}\mbox{ }
\frac{2}{V}\sum_{{\bf{q}}}|M_{{\bf{q}}}|^{2}
 \sin(\omega_{LO}(t-t_{2}))
[exp(i ( \epsilon_{h}({\bf{k-q}})-\epsilon_{h}({\bf{k}}) ) t_{2})
\]
\[
exp(-i ( \epsilon_{h}({\bf{k-q}})+\epsilon_{c}({\bf{k}}) ) t )
\Lambda({\bf{k-q}}t_{2})
\]
\[
 - exp(i ( \epsilon_{c}({\bf{k}})-\epsilon_{c}({\bf{k-q}}) ) t_{2} )
\]
\[
exp(-i ( \epsilon_{c}({\bf{k}})+\epsilon_{h}({\bf{k-q}}) ) t )
\Lambda({\bf{k}}t_{2})]
\Lambda^{*}({\bf{k-q}}t) \Lambda^{*}({\bf{k}}t)
\]
and similarly for $ R_{hh}({\bf{k}}t) $. 
\[
R_{hh}({\bf{k}}t) = -\frac{2i}{V}\sum_{{\bf{q}}}|M_{{\bf{q}}}|^{2}
\sin(\omega_{LO}(t-t_{2}))
[-exp(i(\epsilon_{c}({\bf{k}})-\epsilon_{c}({\bf{k-q}}) t_{2} )
\]
\[
exp(i(\epsilon_{h}({\bf{k}})-\epsilon_{h}({\bf{k-q}}) ) t)
\Lambda({\bf{k}}t_{2})\Lambda^{*}({\bf{k-q}}t)
\]
\[
+exp(i(\epsilon_{h}({\bf{k-q}})-\epsilon_{h}({\bf{k}}))t_{2} )
\]
\[
exp(i(\epsilon_{h}({\bf{k}})-\epsilon_{h}({\bf{k-q}}) )t)
\Lambda({\bf{k-q}}t_{2})\Lambda^{*}({\bf{k-q}}t)]
\]
\[
+ c.c.
\]
 It can be seen that the leading contribution $ R_{he} $ due to phonons
 is proportional to $ \Lambda $ and the next term is proportional to 
 $ \Lambda^{3} $. In the term proportional to $ \Lambda $ it appears that
 we have to use the total vector potential and not the external
 vector potential. This is because corrections to $ \Lambda $  resulting from
 such a replcement is proportional to $ \Lambda^{2} $  which is larger in
 magnitude than the next leding term in $ R_{he} $ which is  proportional to
 $ \Lambda^{3} $. But in fact, this apparent inconsistency is not so severe,
 as we have found that for times when the exciting fields are present,
 the two $ \Lambda $'s are within 1\% of each other. However, for times 
 when the external fields are absent, it is important as a matter
 of principle to include this correction.
 However, we neglect this correction because of a more basic reason, namely
 when the external fields are off the fluctuations in the vector potential
 are going to be comparable to the expectation value of the same.
 Since we are anyway going to treat the fields classically, we are obligated
 to ignore the difference between the two $ \Lambda $ 's as well, after the
 external fields are turned off. 

 The fact the the electron and hole effective masses are different 
 means that $ R_{hh}({\bf{k}}t) $ and $  R_{ee}({\bf{k}}t) $
 are not exactly equal. This is inconsistent with charge conservation.
 A remedy is to use the reduced mass instead of the effective masses
 thereby forcing charge conservation. This approximation will not affect
 the qualitative results and has the added appeal of obeying a
 conservation law.

\subsection{Scaling Relation of the Dephasing Rate with Density at Ultra-Low Densities}

The theory of the semiconductor photon echo is fairly well developed\cite{33}. 
The central feature of the theory is that the energy of the
 echo signal is known to vary exponentially with the the time delay
  $ \tau $ between the two pulses invloved in the echo\cite{16}. Quantitatively,
$ E(\tau)  \mbox{ } \alpha \mbox{ } exp(-\tau/T_{echo}) $ and the
 echo time is related to the polarization dephasing rate $ \gamma $ through
 $ T_{echo} = \frac{1}{4\gamma} $. The echo time can be measured
 by measuring the echo energy as a function of the temporal separation
 of the pulses.
 The polarization dephasing rate $ \gamma $ is a phenomenological
 constant that accounts for the decay of the polarization due to
 various relaxation processes. From our analysis
 it is possible to predict with accuracy the scaling relation of this
 phenomenological dephasing rate with the total carrier density at
 ultra-low densities. We
 have argued that the density dependence of the relaxation rate is
 solely due to carrier-carrier scattering. Our analysis shows that
 $ R_{he}({\bf{k}}t) $ is proportional to $ \Lambda^{3} $ for
 coulomb scattering and proportional to $ \Lambda $ for phonon
 scattering. Further, the total density which is related to
 $ g_{hh}({\bf{k}}t) $ scales as $ \Lambda^{2} $ and $ g_{he}({\bf{k}}t) $
 is proportional to $ \Lambda $ for small $ \Lambda $. Therefore the
 polarization dephasing
 rate at ultra-low densities is proportional to the density (from
 carrier-carrier scattering) and also has
 a added part independent of the density due to phonons.
 A critical
 evaluation (and a measure of the smallness of the density)
 of this conclusion as well as a comparison with the results of other
 authors will be given at the end.

\section{Computational Procedure}
Having obtained the collision terms, we are now ready to solve the 
semiconductor Bloch equations numerically. We follow closely the
approach outlined in Ref.~\onlinecite{31}. We assume an isotropic distribution
and perform all angular integrations analytically. The singular
coulomb terms in the HF part are particularly troublesome
and need special care in their evaluation. This is a well-known
approach outlined in Ref.~\onlinecite{32}. For example, a sum of the type 
shown below can be rewritten as described:
\[
\sum_{{\bf{k \neq k^{'}}}} v_{{\bf{k-k^{'}}}}f(k^{'})
 = \sum_{{\bf{k \neq k^{'}}}} v_{{\bf{k-k^{'}}}}(f(k^{'}) - f(k))
+ f(k)\sum_{{\bf{k \neq k^{'}}}} v_{{\bf{k-k^{'}}}}.
\]
The last part can be evaluated anaytically and in the first part the angular
 intagrations performed analytically. We also assume a cut-off for the
 wavevector $ k_{max} $. The above sum may then be written as
\[
\sum_{{\bf{k \neq k^{'}}}} v_{{\bf{k-k^{'}}}}f(k^{'}) = 
\frac{(4\pi e^{2})}{(2\pi)^{2}}\int_{0}^{k_{max}} dk^{'}\mbox{ }
(\frac{k^{'}}{k})
ln (\left |\frac{(k^{'}+k)}{(k^{'}-k)} \right |)
(f(k^{'}) - f(k))
\]
\[
+ \frac{(4\pi e^{2})}{(2\pi)^{2}}f(k)(\frac{1}{k})
[\frac{1}{2}(k_{max}^{2}-k^{2})
ln (\left |\frac{(k_{max}+k)}{(k_{max}-k)} \right |) + k \mbox{ } k_{max}].
\]
The cutoff $ k_{max} $ is chosen arbitrarily and its validity is justified 
aposteriori. For example, for the case of excitation at the excitonic
 absorption edge we choose
 $ k_{max} = \frac{12.0\pi}{a_{X}} $.
 This choice is justified on the grounds that
 the tail of the occupation probability distribution becomes close to zero
 as the wavevctor approaches $ k_{max} $. We choose 
 $ {\bf{A_{ext}}}(t).{\bf{p_{vc}}} = A_{pump}p_{vc}\cos(\omega_{pump}t) 
\theta(\tau_{X}-t) $
 and $ A_{pump} = \frac{\Lambda_{0}}{(p_{vc} {\frac{e}{mc}} \tau_{X})} $.
 Here, $ p_{vc} $ is the inter-band momentum matrix element,
 $ \tau_{X} = 250fs. $ is the pump duration.
 We can vary the intensity by tuning the
 dimensionless parameter $ \Lambda_{0} $.
 For the numerical solution of the
 semiconductor Bloch equations we use the standard Runge-Kutta
 procedure. The screened coulomb interaction is first numerically
 evaluated and tabulated in an array before proceeding to the Runge-Kutta
 algorithm. It is important to use the rotating wave approximation described
 in detail in the seminal work of R. Binder et. al.\cite{31} in order to get
 sensible results. The collision terms are partly evaluated analytically,
 tabulated and then used in the computations.
 
\section{Results and Discussions}

 The main results of our formalism are shown in Figs.2 and 3.
 The phenomenological
 dephasing rate $ \gamma = 1/T_{2} $ is plotted as a function of the total
 excited carrier density.  The simulations as depicted in Fig.2a confirm our
 claim that at ultra-low density, the dephasing rate scales linearly with
 the density. The scale for the smallness of the density is set by the value
 of $ \Lambda_{0} $. In fact, our simulation shows (Fig.2b) 
 that the total excited
 carrier density at ultra-low densities is given by
 $ \rho_{0} = 7.8\Lambda_{0}^{2} $  (in units of
 $ 10^{15}cm^{-3} $). Therefore, for a choice of $ \Lambda_{0} = 0.1 $
 which is small compared to unity, we find that the upper limit on
 the densities that we can work with is $ 7.8 \times 10^{13}cm^{-3} $
 At these low densities, unfortunately, experimental data are hard to
 come by, because the signal becomes very weak at low densities.
 It is worthwhile to stress the following point. 
 The fact that we are able to extract
 the density dependence of the dephasing rate does not mean that the 
 actual relaxation via carrier-carrier scattering is simply an
 exponential relaxation. The quantity $ T_{2} $
 is merely a useful phenomenological
 parameter that gives us an estimate of the rate of relaxation.
 It also happens to be directly measureable experimentaly,
 via four-wave mixing. Strictly speaking, we should have simulated the
 four-wave mixing experiment including the realistic Coulomb scattering,
 plotted the energy of the echo signal as a function of the time
 delay between the pulses, and then extracted the dephasing rate from 
 the plot. This is much too difficult to carry out and therefore we have
 chosen the simpler route outlined here.

 It is possible to go beyond these densities by making a slight 
 modification to our formalism. The phenomenological collision term 
 $ R_{he} $ contains terms proportional to the powers of the polarization
 via the self-energy and the Green function. The polarization in
 turn decays with a time constant of $ T_{2} $. So, $ R_{he} $ itself
 should decay at the same rate or faster. In our nonequilibrium
 formalism however, this does not happen. The decay has to come from
 the polarization which is proportional to $ \Lambda $.
 This $ \Lambda $ as we have
 defined it, does not decay with time. However, we know that the
 polarization does. So we define a new $ \Lambda $ as follows.
\[
\Lambda^{'}({\bf{k}}t) =  \Lambda({\bf{k}}t)exp(-t/T_{2}) 
\]
 with a $ T_{2} $ to be determined later. Therefore, we replace all the
 $ \Lambda({\bf{k}}t) $ 's in the collision terms by
 $ \Lambda^{'}({\bf{k}}t) $'s. This means that the actual polarization
 (for small $ \Lambda_{0} $)
 is proportional not to $ \Lambda({\bf{k}}t) $  but to
 $ \Lambda^{'}({\bf{k}}t) $. This ensures that the collision terms decay
 with time. The time constant $ T_{2} $ has to be determined self-consistently.
 We vary the value of $ T_{2} $ and plot the total density as a function of
 time(not shown). For time scales that we consider, ( less than 1 $ps$. )
  no recombination
 occurs and so we should see a constant total density when the external
 fields are switched off. For $ \Lambda_{0} $ close to unity and
 large enough $ T_{2} $, we see that the
 theory is asymptotically unstable. In other words, the total density
 increases without saturation. This unphysical result comes about because of
 a lack of self-consistency in the formalism. In order to derive expressions
 for the screened coulomb interaction and the collision terms we have
 ignored the effects of collisions. This inconsistency manifests itself as
 asymptotic instability. In other words, the theory is unreliable at long
 time scales.
 The device of introducing
 $  T_{2} $ is precicesly to partially restore this self-consistency.
 The value of $  T_{2} $ is tuned from a large value to smaller and
 smaller values till we see that the total excited carrier density remains
 constant after the exciting field has been switched off. Any smaller
 value of the $ T_{2} $ than this critical value will also yield the same
 result, of course. However, too small a value of $ T_{2} $ will spoil the
 small time behaviour of our formalism which is modeled exactly.
 Therefore,
 the critical value for $ T_{2} $ provides the dephasing rate. The Fig.3a
 shows the result of the simulations. The experimental data from 
Ref.~\onlinecite{16} 
 shows that at densities larger than $ 10^{16} cm^{-3} $ the polarization
 dephasing rate scales as
 $ \frac{1}{T^{expt.}_{2}} \mbox{ } \alpha \mbox{ } \rho_{0}^{0.33} $ .
 This sublinearity in the scaling is confirmed by our simulations.
 The difference between the predicted exponent and the observed is probably
 atrributable to lack of full self-consistency in the formalism and
 the use of a simple two-band model.
 The sublinearity originates because of a saturation effect on the
 dephasing rate $ \frac{1}{T_{2}} $ brought about by imposing partial
 self-consistency to the formalism.
 We have found that the dephasing rate at relatively high densities
 ($ > 1 \times 10^{15}cm^{-3} $ ) is given by
\[ 
 \frac{1}{T^{theory}_{2}}(in \mbox{ } ps^{-1}) =
 4.0(\rho_{0})^{0.48} 
\]
 where $ \rho_{0} $ is in $ 10^{15}cm^{-3} $.
 In addition, there is a additive contribution of $ 5 \mbox{ } ps^{-1} $
 to $ \frac{1}{T_{2}} $.
 independent of the density (for densities in the range
 $ 10^{12}$ to $5 \times 10^{15} cm^{-3}$) coming from phonons
 (not shown in the figs.).
 Therefore, the carrier-carrier dephasing rate
 becomes equal to the carrier-phonon dephasing rate at a density
 of $ 2 \times 10^{15} cm^{-3} $.
These results are also confirmed by experiments in Refs.~\onlinecite{5} and 
\onlinecite{21}.
 The conventional wisdom that carrier-carrier scattering is much faster than
 carrier-phonon scattering is true at densities much higher than those
 considered here.
 The dependence of the density on the parameter $ \Lambda_{0} $ is shown in
 Fig.3b. The dependence is non-parabolic as $ \Lambda_{0} $ approaches 
 and exceeds unity.
 The dephasing rate for phonons is
 computed in an manner analogous to the method used
 for carrier-carrier scattering. The fact that the dephasing rate is
 independent of the density in the density range considered here suggests
 that the term that is proportional to  $ \Lambda^{3} $ contributes
 insignificantly. We have confirmed this by plotting various quantities
 with and without this term and registering only a marginal change.
 The explanation is, a possible accidental cancellation of the various
 components of this term.
 In fig.4 we see that for a pulse $ \Lambda_{0} = \pi $, in the absence of
 collisions (shown in dotted lines) the phenomenon of quantum beats which is
 the oscillation of the hole occupation probability in time after the
 external field has been turned off ( $ \tau_{X} = 250 fs. $). This is
 due to the superposition of discrete excitonic frequencies resulting in
 the beating behaviour.
 The presence of
 carrier-carrier scattering (in solid) results in the decay of
 the quantum beats with a rate given by $ \frac{1}{T_{2}} = 21.9/ps. $.
 The value of $ \Lambda_{0} = \pi $ is much too large compared to unity
 and significantly beyond the range of applicability of our formalism,
 and therefore the value of $ T_{2} $ quoted above should not be taken
 too literally. The large value of $ \Lambda_{0} $ ensures a large value of
 $ \frac{1}{T_{2}} $ and therefore emphatically illustrates the phenomenon
 of dephasing.
   
\section{Conclusions}

 We have been able to construct an elaborate theory of screening in the
 presense of external classical time varying electromagnetic fields.
 We have obtained explicit formulas for the screened coulomb interaction,
 and demonstrated that they produce finite collision cross-sections
 in both two as well as three dimensions. We have also explained the range
 of validity of these formulas. Using the theory we have
 demonstrated the following:
 The density dependence of the polarization dephasing rate is
 due to carrier-carrier scattering alone, in the range of densities considered,
 and the energy relaxation is due mainly to phonons. Phonons contribute
 to the dephasing rate as an additive term independent of the density.
 We have also
 derived scaling relations of the polarization dephasing rates with density. 
 We have also provided quantitative estimates of the dephasing rates and
 demonstrated that they agree qualitatively with experiments.
 In particular, we have demonsrtated that the dephasing times at
 low densities due to carrier-carrier scattering is in picoseconds and not
 femtoseconds as is sometimes believed.
 We have also studied the
 time evolution of the probability distribution and demonstrated the
 phenomenon of dephasing of excitonic quantum beats as a result of
 carrier-carrier scattering. Further imporovements should involve the 
 use of more realistic band structures such as the inclusion of
 valence subbands. 

\section{Acknowledgements}

	The authors wish to thank Prof. A. J. Leggett for his critical
 review of this work and numerous suggestions and corrections he made
 during the course of this work. We also wish to thank
 Dr. Hanyou Chu for pointing out some of the pitfalls and suggesting
 some remedies during the numerical simulations. This work was
 supported in part by ONR N00014-90-J-1267 and
 the Unversity of Illinois, Materials Research Laboratory under grant
NSF/DMR-89-20539.

\newpage

Fig.1 Figure demontrating the conditions under which the
 simulations are perfeormed.

Fig.2a Linear scaling of the dephasing rate with density at ultra-low density.

Fig.2b Parabolic dependence of the the excited carrier density on
 $ \Lambda_{0} $.

Fig.3a Sublinear scaling of the dephasing rate with density at higher densities.

Fig.3b Non-parabolic dependence of the the excited carrier density on
 $ \Lambda_{0} $ for higher densities.

Fig.4 Dephasing of the excitonic quantum beats as a result of carrier-carrier
 scattering.

\end{document}